\documentclass{report}
\usepackage{setspace,algorithm,algorithmic}
\usepackage{fullpage}
\usepackage{amssymb,amsmath}
\usepackage{epsfig,citesort}

\newcommand{\utwi}[1]{\mbox{\boldmath $ #1$}}

\newcommand{\bu}{{\utwi{u}}}

\newcommand{\bx}{{\utwi{x}}}

\newcommand{\bz}{{\utwi{z}}}

\newcommand {\mm}[1] {\ifmmode{#1}\else{\mbox{\(#1\)}}\fi}
\newcommand{\real} {\mm{{\Bbb R}}}

\DeclareMathOperator{\vol}{vol}
\DeclareMathOperator{\area}{area}

\author{ Jie Liang}
\title{
Computation of protein geometry and its applications: Packing and function prediction
} 
\begin{document}
\maketitle
\tableofcontents

\section{Introduction} 

Three-dimensional atomic structures of protein molecules provide rich
information for understanding how these working molecules of a cell
carry out their biological functions.  With the amount of solved
protein structures rapidly accumulating, computation of geometric
properties of protein structure becomes an indispensable component in
studies of modern biochemistry and molecular biology.  Before we
discuss methods for computing the geometry of protein molecules, we first
briefly describe how protein structures are obtained experimentally.

There are primarily three experimental techniques for obtaining
protein structures: X-ray crystallography, solution nuclear magnetic
resonance (NMR), and recently freeze-sample electron microscopy
(cryo-EM).  In X-ray crystallography, the diffraction patterns of
X-ray irradiation of a high quality crystal of the protein molecule
are measured.  Since the diffraction is due to the scattering of X-ray
by the electrons of the molecules in the crystal, the position, the
intensity, and the phase of each recorded diffraction spot provide
information for the reconstruction of an {\sl electron density map\/}
of atoms in the protein molecule.  Based on independent information of
the amino acid sequence, a model of the protein conformation is then
derived by fitting model conformations of residues to the electron
density map. An iterative process called {\sl refinement} is then
applied to improve the quality of the fit of the electron density
map. The final model of the protein conformation consists of the
coordinates of each of the non-hydrogen atoms \cite{Rhodes-book}.

The solution NMR technique for solving protein structure is based on
measuring the tumbling and vibrating motion of the molecule in
solution. By assessing the chemical shifts of atomic nuclei with spins
due to interactions with other atoms in the vicinity, a set of
estimated distances between specific pairs of atoms can be derived
from NOSEY spectra.  When a large number of such distances are
obtained, one can derive a set of conformations of the protein
molecule, each is consistent with all of the distance constraints
\cite{CrippenHavel88}.  Although determining conformations from either
X-ray diffraction patterns or NMR spectra is equivalent to solving an
ill-posed inverse problem, technique such as Bayesian Markov chain
Monte Carlo with parallel tempering has been shown to be effective in
obtaining protein structures from NMR spectra
\cite{Rieping_Science05}.  The cryo-EM technique for obtaining protein
structure is described in more details in Chapter 11.

\section{Theory and Model}

\subsection{The idealized ball model}
The shape of a protein molecule is complex.  The chemical properties of
atoms in a molecule are determined by their electron charge
distribution.  It is this distribution that generates the scattering
patterns of the X-ray diffraction.  Chemical bonds between atoms lead
to transfer of electronic charges from one atom to another, and
the resulting isosurfaces of the electron density distribution depend
not only on the location of individual nuclei but also on interactions
between atoms. This results in an overall complicated isosurface of
electron density \cite{Bader94}.

The geometric model of macromolecule amenable to convenient
computation is an idealized model, where the shapes of atoms are
approximated by three-dimensional balls. The shape of a protein or a
DNA molecule consisting of many atoms is then the space-filling shape
taken by a set of atom balls.  This model is often called the {\sl
interlocking hard-sphere model\/}, the {\sl fused ball model}, the
{\sl space filling model}
\cite{LeeRichards71,Ri74,Richmond84,Richards85}, or the {\sl union of
ball\/} model \cite{Edels95a}.  In this model, details in the
distribution of electron density, {\it e.g.}, the differences between
regions of covalent bonds and non-covalent bonds, are ignored.  This
idealization is quite reasonable, as it reflects the fact that the
electron density reaches maximum at a nucleus, and its magnitude
decays almost spherically away from the point of the nucleus.  Despite
possible inaccuracy, this idealized model has found wide acceptance,
because it enables quantitative measurement of important geometric
properties (such as area and volume) of molecules.  Insights gained
from these measurements correlate well with experimental observations
\cite{LeeRichards71,Ri77,Richards85,Connolly83_JAC,Richards94_QRB,Gerstein99_ITC}.

In this idealization, the shape of each atom is that of a ball, and
its size parameter is the ball radius.  There are many possible
choices for the parameter set of atomic radii
\cite{Richards74_JMB,Tsai_JMB99}.  Frequently, atomic radii are
assigned the values of their van der Waals radii \cite{Bondi64_JPC}.
Among all these atoms, hydrogen atom has the smallest mass, and has a
much smaller radius than those of other atoms.  For simplification,
the model of {\sl united atom\/} is often employed to approximate the
union of a heavy atom and the hydrogen atoms connected by  a covalent
bond.  In this case, the radius of the heavy atom is increased to
approximate the size of the union of the two atoms.  This practice
significantly reduces the total number of atom balls in the molecule.
However, this approach has been questioned for possible inadequacy
\cite{Word_JMB99}.

The mathematical model of this idealized model is that of the union of
balls \cite{Edels95a}.  For a molecule $M$ of $n$ atoms, the $i$-th
atom is modeled as a ball $b_i$, whose center is located at $\bz_i \in
\real^3$, and the radius of this ball is $r_i \in \real$, namely, we
have $b_i \equiv \{\bx| \bx \in \real^3, ||\bx-\bz_i|| \le r_i \}$
parameterized by $(\bz_i, r_i)$.  The molecule $M$ is formed by the
union of a finite number $n$ of such balls defining the set $\cal B$:
$$
M = \bigcup {\cal B} = \bigcup_{i=1}^n \{b_i\}.
$$ It creates a space-filling body corresponding to the union of the
excluded volumes $\vol( \bigcup_{i=1}^{n} b_i)$ \cite{Edels95a}.  When
the atoms are assigned the van der Waals radii, the boundary surface
$\partial \bigcup {\cal B}$ of the union of balls is called the {\sl van der
Waals} surface.

\subsection{Surface models: Lee-Richards and Connolly's surfaces}
Protein folds into native three-dimensional shape to carry out its
biological functional roles.  The interactions of a protein molecule with other
molecules (such as ligand, substrate, or other protein) 
determine its functional roles.  Such interactions occur physically on
the surfaces of the protein molecule.

\begin{figure}
\centerline{\epsfig{figure=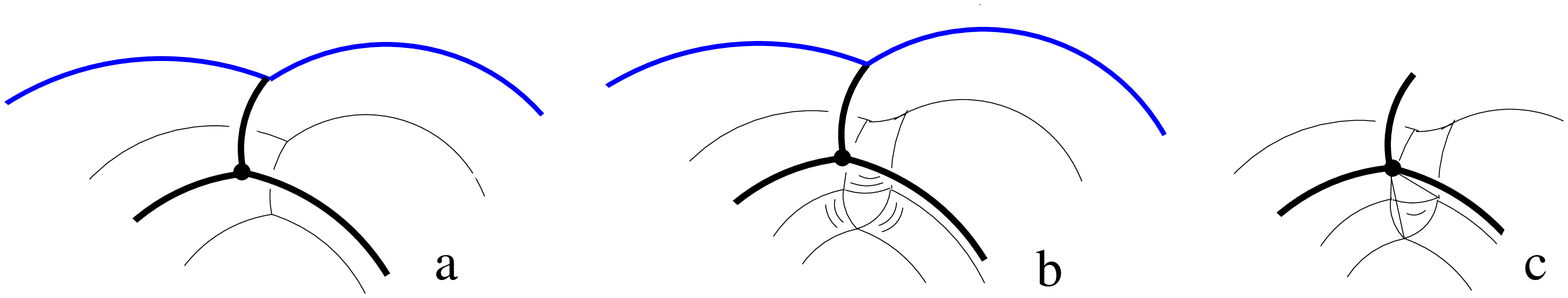,width=5.5in}}
\caption{ Geometric models of protein surfaces.  (a) The solvent
accessible surface (SA surface) is shown in the front.  The van der
Waals surface (beneath the SA surface) can be regarded as a shrunken
version of the SA surface by reducing all atomic radii uniformly by
the amount of the radius of the solvent probe $r_s = 1.4$\AA.  The
elementary pieces of the solvent accessible surface are the three
convex spherical surface pieces, the three arcs, and the vertex where
the three arcs meet.  (b) The molecular surface (MS, beneath the SA
surface) also has three types of elementary pieces: the convex spheric
pieces, which are shrunken version of the corresponding pieces in the
solvent accessible surface, the concave toroidal pieces, and concave
spheric surface.  The latter two are also called the re-entrant
surface. (c) The toroidal surface pieces in the molecular surface,
correspond to the arcs in the solvent accessible surface, and the
concave spheric surface to the vertex.  The set of elements in one
surface can be continuously deformed to the set of elements in the
other surface.}
\label{fig:surface}
\end{figure}

The importance of protein surface was recognized very early on.  Lee
and Richards developed the widely used {\sl solvent accessible
surface} (SA) model, which is also often called the {\sl Lee-Richards
surface} model \cite{LeeRichards71}.  Intuitively, this surface is
obtained by rolling a ball of radius $r_s$ everywhere along the van der
Waals surface of the molecule.  The center of the solvent ball will
then sweep out the solvent accessible surface.  Equivalently, the
solvent accessible surface can be viewed as the boundary surface
$\partial \bigcup {\cal B}_{r_s}$ of the union of a set of inflated balls
${\cal B}_{r_s}$, where each ball takes the position of an atom, but with an
inflated radius $r_i +r_s$ (Fig.~\ref{fig:surface}a).

The solvent accessible surface in general has many sharp crevices and sharp
corners.  In hope of obtaining a smoother surface, one can take
the surface swept out by the front instead of the center of the
solvent ball.  This surface is the {\sl molecular surface} (MS model), which is
often called the {\sl Connolly's surface} after Michael Connolly who
developed the first algorithm for computing molecular surface
\cite{Connolly83_JAC}.  Both solvent accessible surface and molecular 
surface are formed by elementary pieces of simpler shape.

\vspace*{.15 in} \noindent {\bf Elementary pieces.}  For the solvent
accessible surface model, the boundary surface of a molecule consists
of three types of elements: the convex spherical surface pieces, arcs
or curved line segments (possibly a full circle) formed by two
intersecting spheres, and a vertex that is the intersection point of
three atom spheres.  The whole boundary surface of the molecules can
be thought of as a surface formed by stitching these elements
together.

Similarly, the molecular surface swept out by the front of the solvent
ball can also be thought of as being formed by elementary surface
pieces.  In this case, they are the convex spherical surface pieces,
the toroidal surface pieces, and the concave or inverse spherical
surface pieces (Fig.~\ref{fig:surface}b) .  The latter two types of
surface pieces are often called the ``re-entrant surfaces''
\cite{Connolly83_JAC,Richards85}.

The surface elements of the solvent accessible surface and the
molecular surface are closely related.  Imagine a process where atom
balls are shrunk or expanded.  The vertices in solvent accessible
surface becomes the concave spherical surface pieces, the arcs becomes
the toroidal surfaces, and the convex surface pieces become smaller
convex surface pieces (Fig.~\ref{fig:surface}c).  Because of this
mapping, these two type of surfaces are combinatorically equivalent
and have similar topological properties, {\it i.e.}, they are homotopy
equivalent.

However, the SA surface and the MS surface differ in their metric
measurement.  In concave regions of a molecule, often the front of the
solvent ball can sweep out a larger volume than the center of the
solvent ball.  A void of size close to zero in solvent accessible
surface model will correspond to a void of the size of a solvent ball ($4
\pi r_s^3/3$).  It is therefore important to distinguish these two
types of measurement when interpreting the results of volume
calculations of protein molecules.  The intrinsic structures of these
fundamental elementary pieces are closely related to several geometric
constructs we describe below.

\subsection{Geometric constructs}

\begin{figure}
\centerline{\epsfig{figure=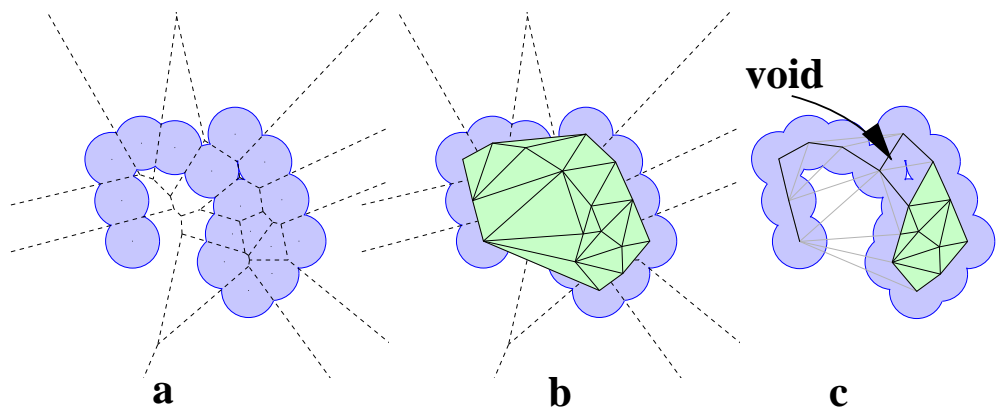,width=5.5in}}
\caption{ Geometry of a simplified two dimensional model molecule, to
illustrate the geometric constructs and the procedure mapping the
Voronoi diagram to the Delaunay triangulation. (a) The molecule formed
by the union of atom disks of uniform size. Voronoi diagram is in
dashed lines. (b) The shape enclosed by the boundary polygon is the
{\sl convex hull}. It is tessellated by the {\sl Delaunay
triangulation}. (c) The alpha shape of the molecule is formed by
removing those Delaunay edges and triangles whose corresponding
Voronoi edges and Voronoi vertices do not intersect with the body of
the molecule.  A molecular void is represented in the alpha shape by
two empty triangles.} \label{fig:vor}
\end{figure}

\vspace*{.15 in} \noindent {\bf Voronoi diagram.}  Voronoi diagram
(Fig~\ref{fig:vor}a), also known as Voronoi tessellation, is a
geometric construct that has been used for analyzing protein packing
in the early days of protein crystallography
\cite{Ri74,Finney75,GeFi82}.  For two dimensional Voronoi diagram, we
consider the following analogy.  Imagine a vast forset containing a
number of fire observation towers.  Each fire ranger is responsible
for putting out any fire closer to his/her tower than to any other
tower.  The set of all trees for which a ranger is responsible
constitutes the Voronoi cell associated with his/her tower, and the
map of ranger responsibilities, with towers and boundaries marked,
constitutes the Voronoi diagram.

We formalize this for three dimensional space.  Consider the point set
$S$ of atom centers in three dimensional space $\real^3$.  The {\sl
Voronoi region\/} or {\sl Voronoi cell\/} $V_i$ of an atom $b_i$ with
atom center $\bz_i \in \real^3$ is the set of all points that are at
least as close to $\bz_i$ than to any other atom centers in ${\cal S}$:
\begin{equation}
V_i = \{ \bx \in \real^3| ||\bx-\bz_i|| \le || \bx - \bz_j||, \bz_j \in
{\cal S} \}.
\end{equation}

We can have an alternative view of the Voronoi cell of an atom $b_i$.
Considering the distance relationship of atom center $\bz_i$ with the
atom center $\bz_k$ of another atom $b_k$. The plane
bisecting the line segment connecting points $\bz_i$ and $\bz_k$
divides the full $\real^3$ space into two half spaces, where points in
one half space is closer to $\bz_i$ than to $\bz_k$,
and points in the other allspice is closer to $\bz_k$ than to
$\bz_i$.  If we repeat this process and take $\bz_k$ in turn from the
set of all atom centers other than $\bz_i$, we will have a number of halfspaces
where points are closer to $\bz_i$ than to each of the atom center
$\bz_k$.  The Voronoi region $V_i$ is then the common intersections of
these half spaces, which is convex.  When we consider
atoms of different radii, we replace the Euclidean distance
$||\bx-\bz_i||$ with the {\sl power distance\/} defined as:
$\pi_i(\bx) \equiv ||\bx-\bz_i||^2 - r_i^2$.

\vspace*{.15 in} \noindent {\bf Delaunay tetrahedrization.}  Delaunay
triangulation in $\real^2$ or Delaunay tetrahedrization in $\real^3$
is a geometric construct that is closely related to the Voronoi
diagram (Fig~\ref{fig:vor}b).  In general, it uniquely tessellates or
tile up the space of the {\sl convex hull\/} of the atom centers in
$\real^3$ with tetrahedra. Convex hull for a point set is the smallest
convex body that contains the point set \footnote{For a two
dimensional toy molecule, we can imagine that we put nails at the
locations of the atom centers, and tightly wrap a rubber band around
these nails. The rubber band will trace out a polygon.  This polygon
and the region enclosed within is the convex hull of the set of points
corresponding to the atom centers.  Similarly, imagine if we can
tightly wrap a tin-foil around a set of points in three dimensional
space, the resulting convex body formed by the tin-foil and space
enclosed within is the convex hull of this set of points in
$\real^3$.}.  The Delaunay tetrahedrization of a molecule can be
obtained from the Voronoi diagram. Consider that the Delaunay
tetrahedrization is formed by gluing four types of primitive elements
together: vertices, edges, triangles, and tetrahedra. Here vertices
are just the atom centers.  We obtain a Delaunay edge by connecting
atom centers $\bz_i$ and $\bz_j$ if and only if the Voronoi regions
$V_i$ and $V_j$ have a common intersection, which is a planar piece
that may be either bounded or extend to infinity.  We obtain a
Delaunay triangle connecting atom centers $\bz_i$, $\bz_j$, and
$\bz_k$ if the common intersection of Voronoi regions $V_i, V_j$ and
$V_k$ exists, which is either a line segment, or a half-line, or a
line in the Voronoi diagram.  We obtain a Delaunay tetrahedra
connecting atom centers $\bz_i,\bz_j,\bz_k$ and $\bz_l$ if and only if
the Voronoi regions $V_i, V_j, V_k$ and $V_l$ intersect at a point.

\subsection{Topological structures}
\vspace*{.15 in} \noindent {\bf Delaunay complex.}  The 
structures in both Voronoi diagram and Delaunay tetrahedrization are
better described with concepts from algebraic topology.  We focus
on the intersection relationship in the Voronoi diagram and introduce
concepts formalizing the primitive elements.  In $\real^3$, between
two to four Voronoi regions may have common intersections.  We use
{\sl simplices\/} of various dimensions to record these intersection
or overlap relationships.  We have vertices $\sigma_0$ as 0-simplices,
edges $\sigma_1$ as 1-simplices, triangles $\sigma_2$ as 2-simplices,
and tetrahedra $\sigma_3$ as 3-simplices.  Each of the Voronoi plane,
Voronoi edge, and Voronoi vertices corresponds to a 1-simplex
(Delaunay edge), 2-simplex (Delaunay triangle), and 3-simplex
(Delaunay tetrahedron), respectively.  If we use 0-simplices to
represent the Voronoi cells, and add them to the simplices
induced by the intersection relationship, we can think of the Delaunay
tetrahedrization as the structure obtained by ``glueing'' these
simplices properly together.  Formally, these simplices form a
{\sl simplicial complex\/} ${\cal K}$:
\begin{equation}
{\cal K} = \{
\sigma_{|I|-1} | \bigcap_{i \in I} V_i \ne \emptyset
\},
\end{equation}
where $I$ is an index set for the vertices representing atoms whose
Voronoi cells overlap, and $|I|-1$ is the dimension of the simplex.

\begin{figure}
\centerline{\epsfig{figure=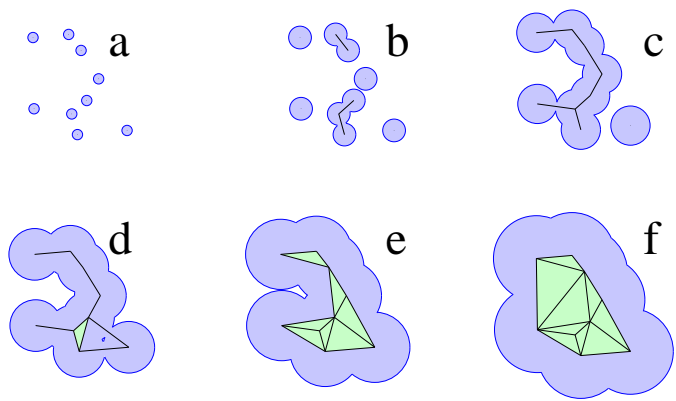,width=4.0in}}
\caption{ The family of alpha shapes or dual simplicial complexes for
a two-dimensional toy molecule.  (a) We collect simplices from the
Delaunay triangulation as atoms grow by increasing the $\alpha$
value. At the beginning as $\alpha$ grows from $-\infty$, atoms are in
isolation and we only have vertices in the alpha shape.  (b) and (c)
When $\alpha$ is increased such that some atom pairs start to
intersect, we collect the corresponding Delaunay edges. (d) When three
atoms intersect as $\alpha$ increases, we collect the corresponding
Delaunay triangles.  When $\alpha = 0$, the collection of vertices,
edges, and triangles form the dual simplicial complex ${\cal K}_0$,
which reflecting the topological structure of the protein
molecule. (e) More edges and triangles from the Delaunay triangulation
are now collected as atoms continue to grow. (d) Finally, all
vertices, edges, and triangles are now collected as atoms are grown to
large enough size.  We get back the full original Delaunay complex.}
\label{fig:filter}
\end{figure}

\vspace*{.15 in} \noindent {\bf Alpha shape and protein surfaces.}
Imagine we can turn a knob to increase or decrease the size of all
atoms simultaneously.  We can then have a model of growing balls and
obtain further information from the Delaunay complex about the shape
of a protein structure.  Formally, we use a parameter $\alpha \in
\real$ to control the size of the atom balls.  For an atom ball $b_i$
of radius $r_i$, we modified its radius $r_i$ at a particular $\alpha$
value to $r_i(\alpha) = (r_i^2+\alpha)^{1/2}$.  When $-r_i<\alpha <
0$, the size of an atom is shrunk. The atom could even disappear if
$\alpha < 0$ and $|\alpha| > r_i$.  We start to collect the simplices
at different $\alpha$ value as we increase $\alpha$ from $-\infty$ to
$+\infty$ (see Fig~\ref{fig:filter} for a two-dimensional example).
At the beginning, we only have vertices.  When $\alpha$ is increased
such that two atoms are close enough to intersect, we collect the
corresponding Delaunay edge that connects these two atom centers.
When three atoms intersect, we collect the corresponding Delaunay
triangle spanning these three atom centers.  When four atoms
intersect, we collect the corresponding Delaunay tetrahedron.  At any
specific $\alpha$ value, we have a {\sl dual simplicial complex\/} or
{\sl alpha complex\/} ${\cal K}_\alpha$ formed by the collected
simplices.  If all atoms take the incremented radius of $r_i + r_s$
and $\alpha=0$, we have the dual simplicial complex ${\cal K}_0$ of
the protein molecule.  When $\alpha$ is sufficiently large, we have
collected all simplices and we get the full Delaunay complex.  This
series of simplicial complexes at different $\alpha$ value form a
family of shapes (Fig~\ref{fig:filter}), called {\sl alpha shapes}, each faithfully
represents the geometric and topological property of the protein
molecule at a particular resolution parametrized by the $\alpha$
value.

An equivalent way to obtain the alpha shape at $\alpha = 0$ is to take
a subset of the simplices, with the requirement that the corresponding
intersections of Voronoi cells must overlap with the body of the union
of the balls. We obtain the dual complex or alpha shape ${\cal K}_0$
of the molecule at $\alpha = 0$ (Fig~\ref{fig:vor}c):
\begin{equation}
{\cal K}_0 = \{
\sigma_{|I|-1} | \bigcap_{i \in I} V_i \cap \bigcup {\cal B}
\ne \emptyset 
\}.
\end{equation}

Alpha shape provides a guide map for computing geometric properties of
the  structures of biomolecules. Take the molecular surface as an example, the
re-entrant surfaces are formed by the concave spherical patch and the
toroidal surface.  These can be mapped from the boundary triangles and
boundary edges of the alpha shape, respectively \cite{Edels95_Hawaii}.
Recall that a triangle in the Delaunay tetrahedrization corresponds to
the intersection of three Voronoi regions, {\it i.e.}, a Voronoi edge.
For a triangle on the boundary of the alpha shape, the corresponding
Voronoi edge intersects with the body of the union of balls by
definition.  In this case, it intersects with the solvent accessible
surface at the common intersecting vertex when the three atoms overlap.
This vertex corresponds to a concave spherical surface patch in the
molecular surface.  For an edge on the boundary of the alpha shape,
the corresponding Voronoi plane coincides with the intersecting plane
when two atoms meet, which intersect with the surface of the union of
balls on an arc.  This line segment corresponds to a
toroidal surface patch.  The remaining part of the surface are convex
pieces, which correspond to the vertices, namely, the atoms on the
boundary of the alpha shape.

The numbers of toroidal pieces and concave spherical pieces are exactly
the numbers of boundary edges and boundary triangles in the alpha
shape, respectively.  Because of the restriction of bond
length and the excluded volume effects, the number of edges and
triangles in molecules are roughly in the order of ${\cal O}(n)$ \cite{Liang98a_Proteins}.

\subsection{Metric measurement}  We have described
the relationship between the simplices and the surface elements of the
molecule.  Based on this type of relationship, we can compute efficiently size
properties of the molecule. We take the problem of volume computation
as an example.

Consider a grossly incorrect way to compute the volume of a protein
molecule using the solvent accessible surface model.  We could define
that the volume of the molecule is the summation of the volumes of
individual atoms, whose radii are inflated to account for solvent
probe.  By doing so we would have significantly inflated the value of
the true volume, because we neglected to consider volume overlaps.  We
can explicitly correct this by following the inclusion-exclusion
formula: when two atoms overlap, we subtract the overlap; when three
atoms overlap, we first subtract the pair overlaps, we then add back
the triple overlap, {\it etc}.  This continues when there are four,
five, or more atoms intersecting.  At the combinatorial level, the
principle of inclusion-exclusion is related to the Gauss-Bonnet
theorem used by Connolly \cite{Connolly83_JAC}.  The corrected volume
$V({\cal B})$ for a set of atom balls ${\cal B}$ can then be written
as:
\begin{equation}
\begin{split}
V({\cal B}) & = 
\sum_{\substack{
\vol(\bigcap T) > 0\\
T \subset {\cal B}
}
} (-1)^{\dim(T)-1} \vol(\bigcap T),
\end{split}
\label{eqn:volume}
\end{equation}
where $\vol(\bigcap T)$ represents volume overlap of various
degree, $T\subset {\cal B}$ is a subset of the balls with non-zero volume
overlap: $\vol(\bigcap T) >0$.

However, the straightforward application of this inclusion-exclusion
formula does not work.  The degree of overlap can be very high:
theoretical and simulation studies showed that the volume overlap can
be up to 7-8 degrees \cite{Kra81,Petitjean94_JCC}.  It is difficult to
keep track of these high degree of volume overlaps correctly during
computation, and it is also difficult to compute the volume of these
overlaps because there are many different combinatorial situations,
{\it i.e.}, to quantify how large is the $k$-volume overlap of which
one of the $7 \choose k$ or $8 \choose k$ overlapping atoms for all of
$k=2, \cdots, 7$ \cite{Petitjean94_JCC}.  It turns out that for
three-dimensional molecules, overlaps of five or more atoms at a time
can always be reduced to a ``$+$'' or a ``$-$'' signed combination of
overlaps of four or fewer atom balls \cite{Edels95a}.  This requires
that the 2-body, 3-body, and 4-body terms in Eqn~\ref{eqn:volume}
enter the formula if and only if the corresponding edge $\sigma_{ij}$
connecting the two balls (1-simplex), triangles $\sigma_{ijk}$
spanning the three balls (2-simplex), and tetrahedron $\sigma_{ijkl}$
cornered on the four balls (3-simplex) all exist in the dual
simplicial complex ${\cal K}_0$ of the molecule
\cite{Edels95a,Liang98a_Proteins}.  Atoms corresponding to these
simplices will all have volume overlaps.  In this case, we have the
simplified exact expansion:
\[
\begin{split}
V({\cal B})
     & =   \sum_{\sigma_i\in {\cal K}} \vol (b_i)  - \sum_{\sigma
_{ij} \in {\cal K}} \vol (b_i \cap b_j)  \\
     & \quad \quad \quad \quad +  \sum _{\sigma_{ijk} \in {\cal K}} \vol 
(b_i \cap b_j \cap b_k)  - \sum _{\sigma_{ijkl}\in {\cal K}} \vol (b_i \cap b_j 
\cap b_k \cap b_l).
\end{split}
\]
The same idea is applicable for the calculation of surface area of
molecules.

\begin{figure}
\centerline{\epsfig{figure=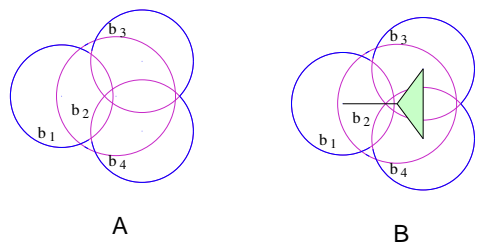,width=3.5in}}
\caption{An example of analytical area calculation.
(a) Area can be computed using the direct inclusion-exclusion.
(b) The formula is simplified without any redundant terms when using alpha shape.
}
\label{fig:pie}
\end{figure}

\noindent {\bf An example.}  An example of area computation by alpha
shape is shown in Fig~\ref{fig:pie}.  Let $b_1, b_2, b_3, b_4$ be the
four disks.  To simplify the notation we write $A_i$ for the area of
$b_i$, $A_{ij}$ for the area of $b_i \cap b_j$, and $A_{ijk}$ for the
area of $b_i \cap b_j \cap b_k$.  The total area of the union, $b_1
\cup b_2 \cup b_3 \cup b_4$, is
\begin{eqnarray*}
  A_{\mbox{total}}  & = &  ( A_1  + A_2  + A_3 + A_4 )  \\
                    & - &  ( A_{12} + A_{23}  + A_{24} + A_{34} ) \\
                    & + &  A_{234} .
\end{eqnarray*}
We add the area of $b_i$ if the corresponding vertex belongs
to the alpha complex (Fig~\ref{fig:pie}),
we subtract the area of $b_i \cap b_j$
if the corresponding edge belongs to the alpha complex,
and we add the area of $b_i \cap b_j \cap b_k$
if the corresponding triangle belongs to the alpha complex.
Note without the guidance of the alpha complex, the
inclusion-exclusion formula may be written as:
 \begin{eqnarray*}
  A_{\mbox{total}}  & = &  ( A_1  + A_2  + A_3 + A_4 )  \\
                    & - &  ( A_{12} + A_{13} + A_{14} + 
                             A_{23}  + A_{24} + 
                             A_{34} ) \\
                    & + &  (A_{123} + A_{124} + A_{134} + A_{234})\\
                    & - &  A_{1234}.
\end{eqnarray*}
This contains 6 canceling redundant terms: $A_{13} = A_{123}$, $A_{14}
= A_{124}$, and $A_{134} = A_{1234}$.  Computing these terms would be
wasteful.  Such redundancy does not occur when we use the alpha
complex: the part of the Voronoi regions contained in the respective
atom balls for the redundant terms do not intersect.  Therefore, the
corresponding edges and triangles do not enter the alpha complex.  In
two dimensions, we have terms of at most three disk intersections,
corresponding to triangles in the alpha complex. Similarly, in three
dimensions the most complicated terms are intersections of four spherical
balls, and they correspond to tetrahedra in the alpha complex.

\vspace*{.15 in} \noindent {\bf Voids and pockets.}  Voids and pockets
represent the concave regions of a protein surface.  Because
shape-complementarity is the basis of many molecular recognition
processes, binding and other activities frequently occur in pocket or
void regions of protein structures.  For example, the majority of
enzyme reactions take place in surface pockets or interior voids.

The topological structure of the alpha shape also offers an effective
method for computing voids and pockets in proteins.  Consider the
Delaunay tetrahedra that are not included in the alpha shape.  If we
repeatedly merge any two such tetrahedra on the condition that they
share a 2-simplex triangle, we will end up with discrete sets of
tetrahedra.  Some of them will be completely isolated from the
outside, and some of them are connected to the outside by triangle(s)
on the boundary of the alpha shape.  The former corresponds to voids
(or cavities) in proteins, the latter corresponds to {\sl pockets\/} and
{\sl depressions\/} in proteins.

\begin{figure}
\centerline{\epsfig{figure=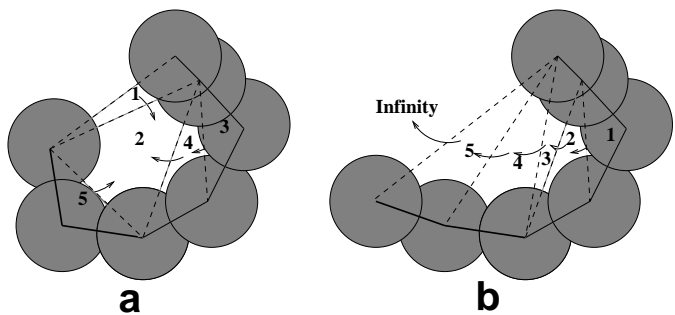,width=3.5in}}
\caption{Discrete flow of empty space illustrated for two dimensional
disks. (a) Discrete flow of a pocket. Triangles 1, 3, 4 and 5 are
obtuse. The free volume flows to the ``sink'' triangle 2, which is
acute. (b) In a depression, the flow is from obtuse triangles to the
outside. }
\label{fig:pockets}
\end{figure}

A pocket differs from a depression in that it must have an opening
that is 
at least narrower than one interior cross-section.  
Formally,
the {\em discrete flow\/} \cite{Edels98_DAM} explains the distinction
between a depression and a pocket.  In a two dimensional Delaunay
triangulation, the empty triangles that are not part of the alpha
shape can be classified into obtuse triangles and acute triangles. The
largest angle of an obtuse triangle is more than 90 degrees, and the
largest angle of an acute triangle is less than 90 degrees. An empty
obtuse triangle can be regarded as a ``source'' of empty space that
``flows'' to its neighbor, and an empty acute triangle a ``sink'' that
collects flow from its obtuse empty neighboring triangle(s). In
Figure~\ref{fig:pockets}a, obtuse triangles 1, 3, 4 and 5 flow to the
acute triangle 2, which is a sink.  Each of the discrete empty spaces
on the surface of protein can be organized by the flow systems of the
corresponding empty triangles: Those that flow together belong to the
same discrete empty space. For a pocket, there is at least one sink
among the empty triangles. For a depression, all triangles are obtuse,
and the discrete flow goes from one obtuse triangle to another, from
the innermost region to outside the convex hull. The discrete flow of
a depression therefore goes to infinity. Figure~\ref{fig:pockets}b
gives an example of a depression formed by a set of obtuse triangles.

Once voids and pockets are identified, we can apply the
inclusion-exclusion principle based on the simplices 
 to compute the exact size measurement ({\it e.g.}, volume
and area) of each void and pocket
\cite{Liang98b_Proteins,Edels98_DAM}.

The distinction between voids and pockets depends on the specific set
of atomic radii and the solvent radius.  When a larger solvent ball is
used, the radii of all atoms will be inflated by a larger amount.
This could lead to two different outcomes.  A void or pocket may
become completely filled and disappear.  On the other hand, the
inflated atoms may not fill the space of a pocket, but may close off
the opening of the pocket.  In this case, a pocket becomes a void.  A
widely used practice in the past was to adjust the solvent ball and
repeatedly compute voids, in the hope that some pockets will become
voids and hence be identified by methods designed for cavity/void
computation.  The pocket algorithm \cite{Edels98_DAM} and tools such
as {\sc CastP} \cite{Liang98_PS,CASTP} often makes this unnecessary.

\section{Computation and software}
\vspace*{.15 in} \noindent {\bf Computing Delaunay tetrahedrization
and Voronoi diagram.}  It is easier to discuss the computation of
tetrahedrization first.  The incremental algorithm developed in
\cite{Edels96_Algorithmica} can be used to compute the weighted
tetrahedrization for a set of atoms of different radii.  For
simplicity, we sketch the outline of the algorithm below for two
dimensional unweighted Delaunay triangulation.

The intuitive idea of the algorithm can be traced back to the original
observation of Delaunay.  For the Delaunay triangulation of a point
set, the circumcircle of an edge and a third point forming a Delaunay
triangle must not contain a fourth point. Delaunay showed that if all
edges in a particular triangulation satisfy this condition, the
triangulation is a Delaunay triangulation.  It is easy to come up with
an arbitrary triangulation for a point set.  A simple algorithm to
covert this triangulation to the Delaunay triangulation is therefore
to go through each of the triangles, and make corrections using
``flips'' discussed below if a specific triangle contains an edge
violating the above condition.  The basic ingredients for computing
Delaunay tetrahedrization are generalizations of these observations.
We discuss the concept of {\sl locally Delaunay\/} edge and the {\sl
edge-flip} primitive operation below.

\begin{figure}
\centerline{\epsfig{figure=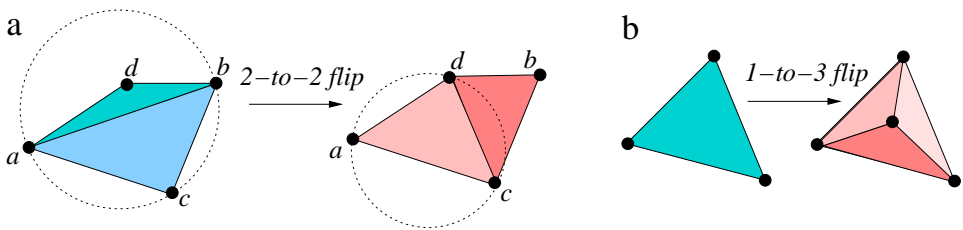,width=5.in}}
\caption{An illustration of {\sl locally Delaunay edge\/} and {\sl
flips}.  (a) For the quadrilateral $abcd$, edge $ab$ is not locally
Delaunay, as the circumcircle passing through edge $ab$ and a third
point $c$ contains a fourth point $d$.  Edge $cd$ is locally Delaunay,
as $b$ is outside the circumcircle $adc$.  An {\sl edge-flip\/} or
{\sl 2-to-2 flip} replaces edge $ab$ by edge $cd$, and replace the
original two triangles $abc$ and $adb$ with two new triangles $acd$
and $bcd$.  (b) When a new vertex is inserted, we replace the old
triangle containing this new vertex with three new triangles.  This is
called {\sl 1-to3\/} flips. }
\label{fig:flip}
\end{figure}

{\it Locally Delaunay edge.} We say an edge $ab$ is locally Delaunay
if either it is on the boundary of the convex hull of the point set,
or if it belongs to two triangles $abc$ and $abd$, and the
circumcircle of $abc$ does not contain $d$ ({\it e.g.}, edge $cd$ in
Fig~\ref{fig:flip}a).

{\it Edge-flip.}  If $ab$ is not locally Delaunay (edge $ab$ in
Fig~\ref{fig:flip}a), then the union of the two triangles $abc \cup
abd$ is a convex quadrangle $acbd$, and edge $cd$ is locally Delaunay.
We can replace edge $ab$ by edge $cd$.  We call this an {\sl
edge-flip} or {\sl 2-to-2 flip}, as two old triangles are replaced by
two new triangles.

We recursively check each boundary edge of the
quadrangle $abcd$ to see if it is also locally Delaunay after
replacing $ab$ by $cd$.  If not, we recursively edge-flip it.

{\it Incremental algorithm for Delaunay triangulation.}  Assume we
have a finite set of points (namely, atom centers) ${\cal S} = \{ \bz_1,
\bz_2, \cdots, \bz_i, \cdots, \bz_n\}$.  We start with a large
auxiliary triangle that contains all these points.  We insert the
points one by one.  At all times, we maintain a Delaunay triangulation
${\cal D}_i$ upto insertion of point $\bz_i$.

After inserting point $\bz_i$, we search for the triangle $\tau_{i-1}$
that contains this new point.  We then add $\bz_i$ to the triangulation
and split the original triangle $\tau_{i-1}$ into three smaller
triangles.  This split is called {\sl 1-to-3 flip}, as it replaces one
old triangle with three new triangles. 
We then check if each of the three
edges in $\tau_{i-1}$ still satisfies the locally Delaunay requirement.
If not, we perform a recursive edge-flip.  This algorithm
is summarized in Algorithm 1.
\begin{algorithm}
\label{alg:tri}
\caption{Delaunay triangulation}
\begin{algorithmic}
\STATE Obtain random ordering of points $\{\bz_1, \cdots, \bz_n\}$;
\FOR{$i=1$ to $n$}
\STATE find $\tau_{i-1}$ such $\bz_i \in \tau_{i-1}$;
\STATE add $\bz_i$, and split $\tau_{i-1}$  into three triangles (1-to-3 flip);
\WHILE{any edge $ab$ not locally Delaunay}
\STATE flip $ab$ to other diagonal $cd$ (2-to-2 edge flip);
\ENDWHILE
\ENDFOR
\end{algorithmic}
\end{algorithm}

In $\real^3$, the algorithm of tetrahedrization becomes more complex,
but the same basic ideas apply.  In this case, we need to locate a
tetrahedron instead of a triangle that contains the newly inserted
point.  The concept of locally Delaunay is replaced by the concept of
{\sl locally convex}, and there are flips different than the 2-to-2
flip in $\real^3$ \cite{Edels96_Algorithmica}.  Although an
incremental approach, {\it i.e.}, sequentially adding points, is not
necessary for Delaunay triangulation in $\real^2$, it is necessary in
$\real^3$ to avoid non-flippable cases and to guarantee that the
algorithm will terminate.  This incremental algorithm has excellent
expected performance \cite{Edels96_Algorithmica}.

The computation of Voronoi diagram is conceptually easy once the
Delaunay triangulation is available.  We can take advantage of the
mathematical duality and compute all of the Voronoi vertices, edges,
and planar faces from the Delaunay tetrahedra, triangles, and edges.
Because one point $\bz_i$ may be an vertex of many Delaunay
tetrahedra, the Voronoi region of $\bz_i$ therefore may contain many
Voronoi vertices, edges, and planar faces. The efficient quad-edge
data structure can be used for software implementation
\cite{GuibasStolfi85}.

\vspace*{.15 in} \noindent {\bf Volume and area computation.}
Let $V$ and $A$ denote the volume and area of the molecule,
respectively, ${\cal K_0}$ for the alpha complex, $\sigma$ for a simplex in
${\cal K}$, $i$ for a vertex, $ij$ for an edge, $ijk$ for a triangle,
and $ijkl$ for a tetrahedron.
The algorithm for volume and area computation can be written as Algorithm~\ref{alg:metric}.
\begin{algorithm}
\caption{Volume and area measurement}
\label{alg:metric}
\begin{algorithmic}
\STATE $V := A := 0.0$;
\FORALL{$\sigma \in {\cal K}$}
\IF{$\sigma$ is a vertex $i$}
\STATE $V := V + \vol(b_i)$; $A := A + \area(b_i)$;
\ENDIF 

\IF{$\sigma$ is an edge $ij$}
\STATE $V := V - \vol(b_i \cap b_j)$; $A := A - \area(b_i \cap b_j)$;
\ENDIF 

\IF{$\sigma$ is a triangle $ijk$}
\STATE $V := V + \vol(b_i \cap b_j \cap b_k)$; $A := A + \area(b_i \cap b_j \cap b_k)$;
\ENDIF 

\IF{$\sigma$ is a tetrahedron $ijkl$}
\STATE $V := V - \vol(b_i \cap b_j \cap b_k \cap b_l)$; $A := A - \area(b_i \cap b_j \cap b_k \cap b_l)$;
\ENDIF 

\ENDFOR
\end{algorithmic}
\end{algorithm}
Additional details of volume and area computation can be found
in \cite{Edels95_Hawaii,Liang98a_Proteins}.

\vspace*{.15 in} \noindent {\bf Software.}  The software package {\sc
Delcx} for computing weighted Delaunay tetrahedrization, {\sc Mkalf}
for computing the alpha shape, {\sc Volbl} for computing volume and
area of both molecules and interior voids and be found at {\tt
www.alphashape.org}.  The {\sc CastP} webserver for pocket computation
can be found at {\tt cast.engr.uic.edu}.  There are other studies that
compute or use Voronoi diagrams of protein structures
\cite{Chakravarty_JBC02,goede97:_voron,Harpaz_Structure94,Harpaz_Structure94},
although not all computes the weighted version which allows atoms to
have different radii.

In this short description of algorithm, we have neglected many details
important for geometric computation. For example, the problem of how
to handle geometric degeneracy, namely, when three points are
co-linear, or when four points are co-planar.  Interested readers
should consult the excellent monograph by Edelsbrunner for a detailed
treatise of these and other important topics in computational geometry
\cite{Edels-mesh}.

\section{Applications: Packing analysis.}

An important application of the Voronoi diagram and volume calculation
is the measurement of protein packing.  Tight packing is an important
feature of protein structure \cite{Ri74,Ri77}, and is thought to play
important roles in protein stability and folding dynamics
\cite{Levitt97_ARB}.  The packing density of a protein is 
measured by the ratio of its van der Waals volume and the volume of the
space it occupies.  One approach is to calculate the packing density
of buried residues and atoms using Voronoi diagram \cite{Ri74,Ri77}.
This approach was also used to derive radii parameters of atoms
\cite{Tsai_JMB99}.

\begin{figure}
\centerline{\epsfig{figure=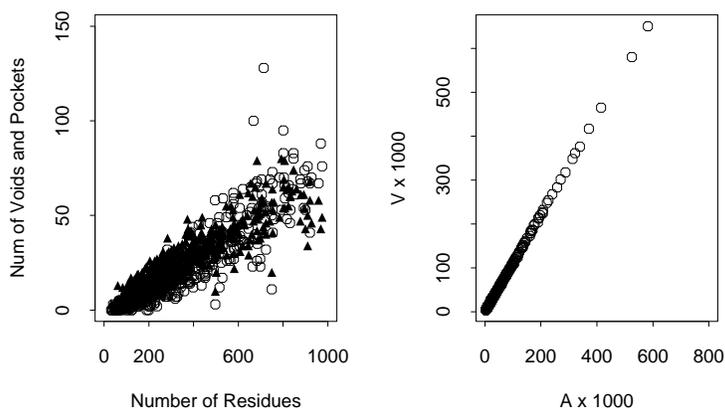,width=4in}}
\caption{ Voids and pockets for a set of 636 proteins representing
most of the known protein folds, and the scaling behavior of the
geometric properties of proteins.  (a) The number of voids and pockets
detected with a 1.4 \AA\ probe is linearly correlated with the number
of residues in a protein. Only proteins with less than 1,000 residues
are shown. Solid triangles and empty circles represent the pockets and
the voids, respectively.  (b) The van der Waals ($vdw$) volume and van
der Waals area of proteins scale linearly with each other.  Similarly,
molecular surface ($ms$) volume also scales linearly with molecular
surface area using a probe radius of 1.4\AA. (Data not shown. Figure
adapted after \cite{LiangDill01_BJ}) }\label{fig:scale}
\end{figure}

Based on the computation of voids and pockets in proteins, a detailed
study surveying major representatives of all known protein structural
folds showed that there is a substantial amount of voids and pockets
in proteins \cite{LiangDill01_BJ}. On average, every 15 residues
introduces a void or a pocket (Fig~\ref{fig:scale}a).  For a perfectly
solid three-dimensional sphere of radius $r$, the relationship between
volume $V = 4\pi r^3/3$ and surface area $A = 4\pi r^2$ is: $V \propto
A^{3/2}$.  In contrast, Figure~\ref{fig:scale}b shows that the van der
Waals volume scales linearly with the van der Waals surface areas of
proteins. The same linear relationship holds irrespective of whether
we relate molecular surface volume and molecular surface area, or
solvent accessible volume and solvent accessible surface area.  This
and other scaling behavior point out that protein interior is not
packed as tight as solid \cite{LiangDill01_BJ}. Rather, packing
defects in the form of voids and pockets are common in proteins.

If voids and pockets are prevalent in proteins, an interesting
question is what is then the origin of the existence of these voids
and pockets.  This question was studied by examining the scaling
behavior of packing density and coordination number of residues
through the computation of voids, pockets, and edge simplices in the
alpha shapes of random compact chain polymers \cite{ZCTL03-jcp}.  For
this purpose, a 32-state discrete state model was used to generate a
large ensemble of compact self-avoiding walks.  This is a difficult
task, as it is very challenging to generate a large number of
independent conformations of very compact chains that are
self-avoiding.  The results in \cite{ZCTL03-jcp} showed that it is
easy for compact random chain polymers to have similar scaling
behavior of packing density and coordination number with chain length.
This suggests that proteins are not optimized by evolution to
eliminate voids and pockets, and the existence of many pockets and
voids is random in nature, and is due to the generic requirement of
compact chain polymers.  The frequent occurrence and the origin of
voids and pockets in protein structures raise a challenging question:
How can we distinguish voids and pockets that perform biological
functions such as binding from those formed by random chance?  This
question is related to the general problem of protein function prediction.

\section{Applications: Protein function prediction from structures.}
Conservation of protein structures often reveals very distant
evolutionary relationship, which are otherwise difficult to detect by
sequence analysis \cite{Todd01_JMB}.  Comparing protein structures can
provide insightful ideas about the biochemical functions of proteins
({\it e.g.}, active sites, catalytic residues, and substrate
interactions)
\cite{Holm97_Structure,Martin98_Structure,Orengo99_COSB}.

A fundamental challenge in inferring protein function from structure
is that the functional surface of a protein often involves only a
small number of key residues.  These interacting residues are
dispersed in diverse regions of the primary sequences and are
difficult to detect if the only information available is the primary
sequence.  Discovery of local spatial motifs from structures that are
functionally relevant has been the focus of many studies.

\vspace*{.15 in} \noindent {\bf Graph based methods for spatial patterns in proteins.}
To analyze local spatial patterns in proteins.  Artymiuk {\it et al}
developed an algorithm based on subgraph isomorphism detection
\cite{Artymiuk94_JMB}.  By representing residue side-chains as
simplified pseudo-atoms, a molecular graph is constructed to represent
the patterns of side-chain pseudo-atoms and their inter-atomic
distances.  A user defined query pattern can then be searched rapidly
against the Protein Data Bank for similarity relationship.  Another
widely used approach is the method of geometric hashing.  By examining
spatial patterns of atoms, Fischer {\it et al} developed an algorithm
that can detect surface similarity of proteins
\cite{Fischer93_Proteins,Norel94_PE}.  This method has also been
applied by Wallace {\it et al} for the derivation and matching of
spatial templates \cite{Wallace97_PS}.  Russell developed a different
algorithm that detects side-chain geometric patterns common to two
protein structures \cite{Russell98_JMB_sidechain}.  With the
evaluation of statistical significance of measured root mean square
distance, several new examples of convergent evolution were
discovered, where common patterns of side-chains were found to reside
on different tertiary folds.

These methods have a number of limitations.  Most require a
user-defined template motif, restricting their utility for automated
database-wide search.  In addition, the size of the spatial pattern
related to protein function is also often restricted.

\vspace*{.15 in} \noindent {\bf Predicting protein functions by
matching pocket surfaces.}  Protein functional surfaces are frequently
associated with surface regions of prominent concavity
\cite{Laskowski96_PS,Liang98_PS}.  These include pockets and voids,
which can be accurately computed as we have discussed.  
Computationally, one wishes to automatically identify
voids and pockets on protein structures where interactions exist with
other molecules such as substrate, ions, ligands, or other proteins.

Binkowski {\it et al}.\ developed a method for predicting protein
function by matching a surface pocket or void on a protein of unknown
or undetermined function to the pocket or void of a protein of known
function \cite{pvsoar03,Binkowski_PS05}.  Initially, the Delaunay tetrahedrization
and alpha shapes for almost all of the structures in the PDB databank
are computed \cite{CASTP}.  All surface pockets and interior voids for
each of the protein structure are then exhaustively computed
\cite{Edels98_DAM,Liang98b_Proteins}.  For each pocket and void, the
residues forming the wall are then concatenated to form a short
sequence fragment of amino acid residues, while ignoring all
intervening residues that do not participate in the formation of the
wall of the pocket or void.  Two sequence fragments, one from the
query protein and another from one of the proteins in the database,
both derived from pocket or void surface residues, are then compared using
dynamic programming.  The similarity score for any observed match is
assessed for statistical significance using an empirical randomization
model constructed for short sequence patterns.

For promising matches of pocket/void surfaces showing significant
sequence similarity, we can further evaluate their similarity in shape
and in relative orientation.  The former can be obtained by measuring
the coordinate root mean square distance ({rmsd}) between the two
surfaces.  The latter is measured by first placing a unit sphere at
the geometric center $\bz_0 \in \real^3$ of a pocket/void.  The
location of each residue $\bz = (x, y, z)^T$ is then projected onto
the unit sphere along the direction of the vector from the geometric
center: $\bu = (\bz-\bz_0)/||\bz -\bz_0 ||$.  The projected pocket is
represented by a collection of unit vectors located on the unit
sphere, and the original orientation of residues in the pocket is
preserved.  The {\sc rmsd} distance of the two sets of unit vectors
derived from the two pockets are then measured, which is called the
o{\sc rmsd} for {\sl orientation {\sc rmsd}} \cite{pvsoar03}.  This
allows similar pockets with only minor conformational changes to be
detected \cite{pvsoar03}.

The advantage of the method of Binkowski {\it et al\/} is that it does
not assume prior knowledge of functional site residues, and does not
require {\it a priori\/} any similarity in either the full primary
sequence or the backbone fold structures.  It has no limitation in the
size of the spatially derived motif and can successfully detect
patterns small and large.  This method has been successfully applied
to detect similar functional surfaces among proteins of the same fold
but low sequence identities, and among proteins of different fold
\cite{pvsoar03,PVSOAR04}.

\vspace*{.15 in} \noindent {\bf Function prediction through models of
protein surface evolution.}  To match local surfaces such as pockets
and voids and to assess their sequence similarity, an effective scoring matrix
is critically important.  In the original study of Binkowski {\it et
al}, {\sc Blosum} matrix was used.  However, this is problematic, as
{\sc Blosum} matrices were derived from analysis of precomputed large
quantities of sequences, while the information of the particular
protein of interest has limited or no influence.  In addition, these
precomputed sequences include buried residues in protein core, whose
conservation reflects the need to maintain protein stability rather
than to maintain protein function.  In reference
\cite{TsengLiang05-EMBC,TsengLiang05-MBE}, a continuous time Markov
process was developed to explicitly model the substitution rates of
residues in binding pockets.  Using a Bayesian Markov chain Monte
Carlo method, the residue substitution rates at functional pocket are
estimated.  The substitution rates are found to be
very different for residues in the binding site and residues on the
remaining surface of proteins.  In addition, substitution rates are
also very different for residues in the buried core and residues on
the solvent exposed surfaces.

These rates are then used to generate a set of scoring
matrices of different time intervals for residues located in the
functional pocket.  Application of protein-specific and
region-specific scoring matrices in matching protein surfaces result
in significantly improved sensitivity and specificity in protein
function prediction \cite{TsengLiang05-EMBC,TsengLiang05-MBE}.

\begin{figure}[t]
\centerline{\epsfig{figure=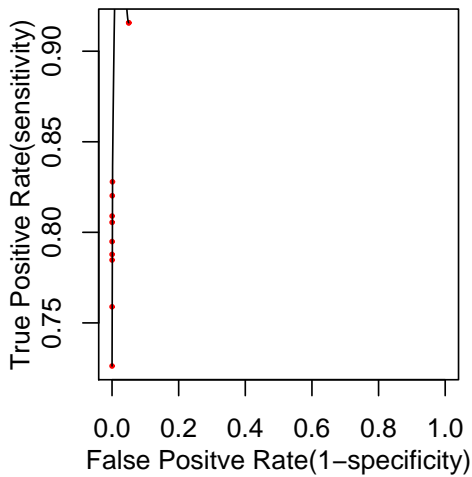,width=2.5in}}
\caption{A large scale study of protein function prediction from
structures by matching similar functional surfaces for 100 protein
families.  A correct prediction is made if the matched surface comes
from a protein structure with the same Enzyme Classification (E.C.)
number (upto the 4-th digit) as that of the query protein.  The
$x$-axis of the Receiver Operating Characteristics curve reflects the
false positive rate ($1-$specificity) at different statistical
significance $p$-value by cRMSD measurement, and the $y$-axis reflects
the true positive rate (sensitivity).}
\label{fig:roc}
\end{figure}

\begin{figure}[!t]
\epsfxsize=3.in   
{\center \epsfbox{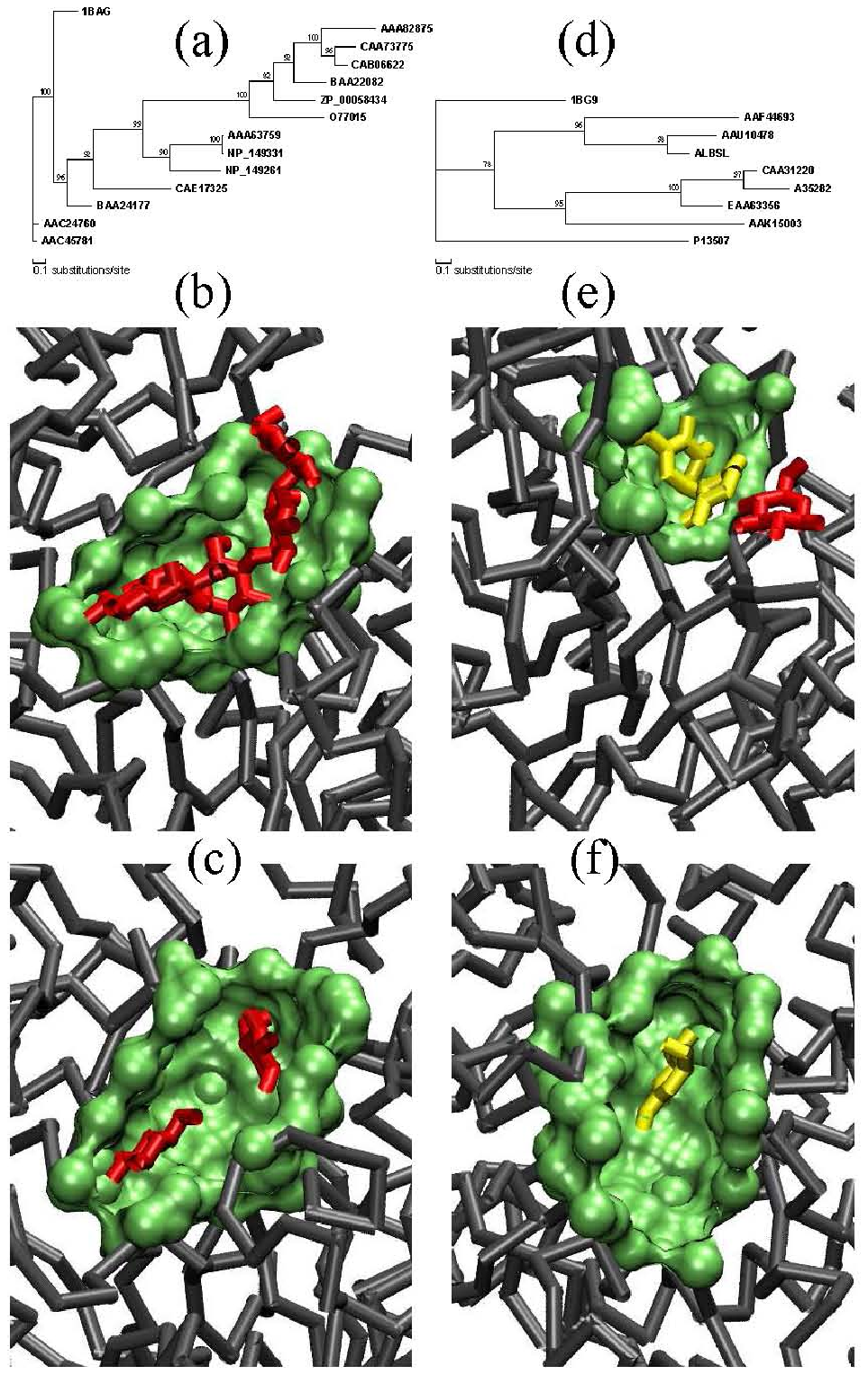}}
\caption{Protein function prediction as illustrated by the example of alpha amylases. Two template binding surfaces are used to search database of protein surfaces to identify
protein structures that are of similar functions.
(a) The phylogenetic tree for the template {\sc Pdb} structure {\tt 1bag} from {\it B.\ subtilis}.
(b) The template binding pocket of alpha amylase on {\tt 1bag}.
(c) A matched binding surface on a different protein structure ({\tt 1b2y} from
human, full sequence identity 22\%)
obtained by
querying with {\tt 1bag}.
(d) The phylogenetic tree for the template structure {\tt 1bg9} from {\it H.\ vulgare}.
(e) The template binding pocket on {\tt 1bg9}.
(f) A matched binding surface on a different protein structure
({\tt 1u2y} from human, full sequence identity 23\%)
obtained by
querying with {\tt 1bg9} (Adapted from  \cite{TsengLiang05-MBE}).
}
\label{fig:amylase}
\vspace *{-5 pt}
\end{figure}

In a large scale study of predicting protein functions from
structures, a subset of 100 enzyme families are collected from the
total of 286 enzyme families containing between 10--50 member protein
structures with known Enzyme Classification (E.C.)\ labels.  By
estimating the substitution rate matrix for residues on the active
site pocket of a query protein, a series of scoring matrices of
different evolutionary time is derived.  By searching for similar
pocket surfaces from a database of 770,466 pockets derived 
from the {\sc CastP} database (with the criterion that each
must contain at least 8 residues), this method can recover active site
surfaces on enzymes similar to that on the query structure at an
accuracy of $>92$\%. Fig~\ref{fig:roc} shows the Receiver Operating
Characteristic Curve of this study. An example of identifying human
amylase using template surfaces from {\it B.\ subtilis} and from
barley is shown in Fig~\ref{fig:amylase}.

The method of surface matching based on evolutionary model is also
especially effective in solving the challenging problems of protein
function prediction of orphan structures of unknown function (such as
those obtained in structural genomics projects), which have only
sequence homologs that are themselves hypothetical proteins with
unknown functions.

\section{Discussion}
A major challenge in studying protein geometry is to understand our
intuitive notions of various geometric aspects of molecular shapes,
and to quantify these notions with mathematical models that are
amenable to fast computation.  The advent of the union of ball model
of protein structures enabled rigorous definition of important
geometric concepts such as solvent accessible surface and molecular
surface.  It also led to the development of algorithms for area and
volume calculations of proteins.  Deep understanding of the
topological structure of molecular shapes is also based on the
idealized union of ball model \cite{Edels95a}.  A success in
approaching these problems is exemplified in the development of the
pocket algorithm \cite{Edels98_DAM}.  Another example is the recent
development of a rigorous definition of protein-protein binding or interaction interface and
algorithm for its computation \cite{Ban04-RECOMB}.

Perhaps a more fundamental problem we face is to identify important
structural and chemical features that are the determinants of
biological problems of interest. For example, we would like to know
what are the shape features that has significant influences on protein solvation, protein
stability, ligand specific binding, and protein conformational
changes.  It is not clear whether our current geometric intuitions are
sufficient, or are the correct or the most relevant ones. There may
still be important unknown shape properties of molecules that elude us
at the moment.

An important application of geometric computation of protein
structures is to detect patterns important for protein function.  The
shape of local surface regions on a protein structure and their
chemical texture are the basis of its binding interactions with other
molecules.  Proteins fold into specific native structure to form these
local regions for carrying out various biochemical functions.  The
geometric shape and chemical pattern of the local surface regions, and
how they change dynamically are therefore of fundamental importance in
computational studies of proteins.

Another important application is the development of geometric
potential functions.  Potential functions are important for generating
conformations, for distinguishing native and near native conformations
from other decoy conformations in protein structure predictions
\cite{Singh96_JCB,Zheng97,Li&Liang03_Proteins,LiLiang_Proteins05} and in protein-protein
docking \cite{LiLiang_PSB05}.  They are also important for peptide and
protein design \cite{LiLiang_PSB05,HuLiLiang04_Bioinformatics}.
Chapter 4 describes in details the development of geometric potential
and applications in decoy discrimination and in protein-protein
docking prediction.

We have not described in detail the approach of studying protein
geometry using graph theory.  In addition to side-chain pattern
analysis briefly discussed earlier, graph based protein geometric model also
has lead to a number of important insights, including the optimal
design of model proteins formed by hydrophobic and polar residues
\cite{Kleinberg99-RECOMB}, and methods for optimal design of
side-chain packing \cite{Xu05-RECOMB,Snoeyink05-PSB}.  Another
important topic we did not touch upon is the analysis of the topology
of protein backbones.  Based on concepts from knot theory, R{\o}gen
and Bohr developed a family of global geometric measures for protein
structure classification \cite{Rogen_MB03}. These measures originate
from integral formulas of Vassiliev knot invariants.  With these
measures, R{\o}gen and Fain further constructed a system that can
automatically classify protein chains into folds
\cite{Rogen_PNASUSA03}.  This system can reproduce the {\sc Cath}
classification system that requires explicit structural alignment as
well as human curation.

Further development of descriptions of geometric shape and topological
structure, as well as algorithms for their computation will provide a solid
foundation for studying many important biological problems.  The other
important tasks are then to show how these descriptors may be
effectively used to deepen our biological insights and to develop
accurate predictive models of biological phenomena.  For example, in
computing protein-protein interfaces, a challenging task is to
discriminate surfaces that are involved in protein binding from other
non-binding surface regions, and to understand in what fashion this
depends on the properties of the binding partner protein.

Undoubtedly, evolution plays central roles in shaping up the function
and stability of protein molecules. The method of analyzing residue
substitution rates using a continuous time Markov models
\cite{TsengLiang05-EMBC,TsengLiang05-MBE}, and the method of surface mapping of
conservation entropy and phylogeny
\cite{Lichtarge_JMB96,Glaser_Bioinformatics03} only scratches the
surface of this important issue.  Much remains to be done in
incorporating evolutionary information in protein shape analysis for
understanding biological functions.

\section{Summary} 
The accumulation of experimentally solved molecular structures of
proteins provides a wealth of information for studying many important
biological problems.  With the development of a rigorous model of the
structure of protein molecules, various shape properties, including
surfaces, voids, and pockets, and measurements of their metric
properties can be computed.  Geometric algorithms have found important
applications in protein packing analysis, in developing potential
functions, in docking, and in protein function prediction.  It is likely further
development of geometric models and algorithms will find important
applications in answering additional biological questions.

\section{Further reading} 
The original work of Lee and Richards surface can be found in
\cite{LeeRichards71}, where they also formulated the molecular surface
model \cite{Richards85}.  Michael Connolly developed the first method
for the computation of the molecular surface \cite{Connolly83_JAC}.
Tsai {\it et al}.\ described a method for obtaining atomic radii
parameter \cite{Tsai_JMB99}.  The mathematical theory of the union of
balls and alpha shape was developed by Herbert Edelsbrunner and
colleague \cite{Edels95a,Edels94_ACMTG}.  Algorithm for computing
weighted Delaunay tetrahedrization can be found in
\cite{Edels96_Algorithmica}, or in a concise monograph with in-depth
discussion of geometric computing \cite{Edels-mesh}.  Details of area
and volume calculations can be found in
\cite{Edels95_Hawaii,Liang98a_Proteins,Liang98b_Proteins}.  The theory
of pocket computation and applications can be found in
\cite{Edels98_DAM,Liang98_PS}.  Richards and Lim offered a
comprehensive review on protein packing and protein folding
\cite{Richards94_QRB}. A detailed packing analysis of proteins can be
found in \cite{LiangDill01_BJ}.  The study on inferring protein
function by matching surfaces is described in \cite{pvsoar03}.  The
study of the evolutionary model of protein binding pocket and its
application in protein function prediction can be found in
\cite{TsengLiang05-MBE}.

\section{Acknowledgments} 
This work is supported by grants from the National Science Foundation
(CAREER DBI0133856), the National Institute of Health (GM68958), the
Office of Naval Research (N000140310329), and the Whitaker Foundation
(TF-04-0023).  The author thanks Jeffrey Tseng for help in preparing
this chapter.

\end{document}